\documentclass[pre,twocolumn,showpacs,superscriptaddress]{revtex4}

\usepackage{graphicx}
\usepackage{color}
\usepackage{amsmath,amssymb}
\usepackage{mathrsfs}

\begin{document}

\title{Aging Scaled Brownian Motion}

\author{Hadiseh Safdari}
\affiliation{Department of Physics, Shahid Beheshti University, G.C., Evin,
Tehran 19839, Iran}
\affiliation{Institute of Physics \& Astronomy, University of Potsdam, 14476
Potsdam-Golm, Germany}
\author{Aleksei V. Chechkin}
\affiliation{Institute for Theoretical Physics, Kharkov Institute of Physics
and Technology, Kharkov 61108, Ukraine}
\affiliation{Institute of Physics \& Astronomy, University of Potsdam, 14476
Potsdam-Golm, Germany}
\affiliation{Max-Planck Institute for the Physics of Complex Systems,
01187 Dresden, Germany}
\author{Gholamreza R. Jafari}
\affiliation{Department of Physics, Shahid Beheshti University, G.C., Evin,
Tehran 19839, Iran}
\author{Ralf Metzler}
\email{rmetzler@uni-potsdam.de}
\affiliation{Institute of Physics \& Astronomy, University of Potsdam, 14476
Potsdam-Golm, Germany}
\affiliation{Department of Physics, Tampere University of Technology, FI-33101
Tampere, Finland}

\date{\today}

\begin{abstract}
Scaled Brownian motion (SBM) is widely used to model anomalous diffusion of
passive tracers in complex and biological systems. It is a highly non-stationary
process governed by the Langevin equation for Brownian motion, however, with
a power-law time dependence of the noise strength. Here we study the aging
properties of SBM for both unconfined and confined motion. Specifically, we
derive the ensemble and time averaged mean squared displacements and analyze
their behavior in the regimes of weak, intermediate, and strong aging. A
very rich behavior is revealed for confined aging SBM depending on different
aging times and whether the process is sub- or superdiffusive. We
demonstrate that the information on the aging factorizes with respect to the
lag time and exhibits a functional form, that is identical to the aging
behavior of scale free continuous time random walk processes. While SBM
exhibits a disparity between ensemble and time averaged observables and is
thus weakly non-ergodic, strong aging is shown to effect a convergence of the
ensemble and time averaged mean squared displacement. Finally, we derive the
density of first passage times in the semi-infinite domain that features a
crossover defined by the aging time.
\end{abstract}


\maketitle

\section{Introduction}
\label{sec:1}

Deviations from normal Brownian motion were reported already in the work of
Richardson on the spreading of tracers in turbulent flows \cite{richardson},
and deviations from the Brownian law are discussed by Freundlich and Kr{\"u}ger
\cite{freundlich}. Today \emph{anomalous diffusion\/} is typically defined in
terms of the power-law form
\begin{equation}
\label{msd}
\langle x^2(t)\rangle\sim2K_{\alpha}^*t^{\alpha}
\end{equation} 
of the mean squared displacement (MSD) \cite{report,bouchaud}. Depending on
the value of the anomalous diffusion exponent we distinguish the regimes of
subdiffusion ($0<\alpha<1$) and superdiffusion ($\alpha>1$), including the
special cases of Brownian motion ($\alpha=1$) and ballistic transport ($\alpha
=2$). The generalized diffusion coefficient $K_{\alpha}^*$ in Eq.~(\ref{msd})
has the physical dimension $\mathrm{cm}^2/\mathrm{sec}^{\alpha}$.

Anomalous diffusion is observed in a wide range of systems, including fields as
diverse as charge carrier motion in amorphous and polymeric semiconductors
\cite{scher,schubert}, dispersion of chemicals in groundwater aquifers
\cite{scher1}, particle dispersion in colloidal glasses \cite{weeks}, or the 
motion of tracers in weakly chaotic systems \cite{solomon}. With the rise of
experimental techniques such as fluorescence correlation spectroscopy or
advanced single particle tracking methods, the discovery of anomalous diffusion has
gone through a sharp rise for the motion of endogenous and artificial tracers in
living biological cells \cite{hoefling,saxton,lene,tabei,weber,bronstein}.
Concurrent to this development an increasing amount of anomalous diffusion
studies is reported in artificially crowded environments mimicking aspects
of the superdense state of the cellular fluid \cite{pan,lene1}. Within
and along lipid membranes anomalous diffusion was found from experiment and
simulations \cite{membranes}.

Brownian motion is intimately connected with the Gaussian probability density
function describing the spatial spreading of a test particle as function of
time. This Gaussian is effected \emph{a forteriori\/} by the central limit
theorem, as Brownian motion is well described on a stochastic level by
the Wiener process. Anomalous diffusion loses this
universal character, and instead different scenarios corresponding to the
physical setting need to be considered. Among the most popular models we
mention the Scher-Montroll continuous time random walk (CTRW), in which
individual jumps are separated by independent, random waiting times
\cite{scher,montroll}. If the distribution of these waiting times is
scale free, subdiffusion emerges \cite{klablushle}. Fractional Brownian
motion and the closely related fractional Langevin equation motion are
stochastic processes fueled by Gaussian yet power-law correlated noise
\cite{mandelbrot,lutz}. Anomalous diffusion emerges when a conventional
random walker is confined to move on a matrix with a fractal dimension
\cite{klemm,fractal,saxton}. Stochastic processes with multiplicative
noise, corresponding to a space-dependent diffusion coefficient, also effect
anomalous diffusion \cite{hdp,fulinsky}. A contemporary summary of different
anomalous diffusion processes exceeding the scope of this introduction is
provided in Ref.~\cite{pccp}.

Here we deal with the remaining of these popular anomalous diffusion models,
namely Scaled Brownian Motion (SBM). SBM is a highly non-stationary process
defined in terms of the stochastic equation
\begin{eqnarray}
\label{langevin}
\frac{dx(t)}{dt}=\sqrt{2\mathscr{K}(t)}\times\xi(t),
\end{eqnarray}
which is driven by white Gaussian noise of zero mean $\langle\xi(t)\rangle=0$
and with autocorrelation $\langle\xi(t_1)\xi(t_2)\rangle=\delta(t_1-t_2)$ .
The explicitly time dependent diffusion coefficient is taken as
\begin{equation}
\label{time_diff}
\mathscr{K}(t)=\alpha K_{\alpha}^*t^{\alpha-1},
\end{equation}
We allow $\alpha$ to range in the interval $(0,2)$,
such that the process describes both subdiffusion and sub-ballistic
superdiffusion. The idea of a power-law time dependent diffusion coefficient
is essentially dating back to Batchelor (albeit he used $\alpha=3$) in his
approach to Richardson turbulent diffusion \cite{batchelor}. SBM, especially
in its subdiffusive form, is widely used to describe anomalous diffusion
\cite{saxton1}. SBM was studied systematically in Refs.~\cite{lim,fulinsky,
thiel,sbm}.

In stationary systems correlations measured between two times $t_1$ and $t_2$
are typically solely functions of the time difference, $f(|t_1-t_2|)$. In
non-stationary systems this functional dependence is generally more involved,
e.g., it can acquire the form $f(t_2/t_1)$ \cite{stas}. In such a non-stationary
setting the origin of time can no longer be chosen arbitrarily. This raises the
question of aging, that is, the explicit dependence of physical observables on
the time span $t_a$ between the original preparation of the system and the start
of the recording of data. Traditionally, aging is considered a key property of
glassy systems \cite{donth}. The aging time $t_a$ can be adjusted deliberately
in certain experiments, such as for the time of flight measurements of charge
carriers in polymeric semiconductors in which the system is prepared by knocking
out the charge carriers by a light flash \cite{schubert}. Similarly, aging could
be checked directly in blinking quantum dot systems, in which the initiation time
is given by the first exposure of the quantum dot to the laser light source. In
other systems, for instance, the motion of tracers in living biological cells,
the aging time is not always precisely defined. In such cases it is therefore
important to have cognisance of the functional effects of aging as developed here.

In the following, we will analyze in detail the aging properties encoded in the
SBM dynamics in both unconfined and confined settings. For free aging SBM in
Section \ref{sec:2} we show that the result for the time averaged MSD factorizes
into a term containing all the information on the aging time $t_a$ and another
capturing the physically relevant dependence on the lag time $\Delta$. This
factorization is identical to that of heterogeneous diffusion processes and
scale-free, subdiffusive CTRW processes.
In Section \ref{sec:3} we explore the aging dynamics of
confined SBM. For increasing aging time $t_a$ it is demonstrated that the
non-stationary features of SBM under confinement are progressively washed out,
a feature, which is important for the evaluation of measured time series.
Section \ref{sec:4} reports the first passage time density on a semi-infinite
domain for aged SBM which includes a crossover between two scaling regimes as a
result of the additional time scale introduced by $t_a$. Finally, Sec.~\ref{sec:5}
concludes this paper.

\section{Ageing effect on unconfined SBM}
\label{sec:2}

The position autocorrelation function (covariance) for SBM in the conventional
(ensemble) sense reads \cite{lim}
\begin{eqnarray}
\label{5a1}
\left\langle x(t_1)x(t_2)\right\rangle=2K_{\alpha}^*\min(t_1,t_2)^{\alpha}.
\end{eqnarray}
For an aged process, in which we measure the MSD starting from the aging time
$t_a$ until time $t$, the result for the MSD thus becomes
\begin{eqnarray}
\label{5a}
\nonumber
\langle x^2(t)\rangle_a&=&\langle[x(t_a+t)-x(t_a)]^2\rangle\\
&=&2K_{\alpha}^*[(t+t_a)^{\alpha}-t_a^{\alpha}].
\end{eqnarray}
For a non-aged process with $t_{a}=0$ the standard scaling (\ref{msd}) of the MSD
is recovered, as it should. In the aged process, the MSD (\ref{5a}) is reduced by
the amount accumulated until time $t_a$, at which the measurement starts. The
limiting cases of expression (\ref{5a}) interestingly reveal the crossover behavior
\begin{equation}
\langle x^2(t)\rangle_a=\left\{\begin{array}{ll}2\alpha K_{\alpha}^*t_a^{\alpha-1}
t, & t_a\gg t\\[0.2cm]
2K_{\alpha}^*t^{\alpha}, & t\gg t_a
\end{array}\right..
\label{6}
\end{equation}
While for weak aging ($t_a\ll t$) the aged MSD (\ref{5a}) becomes identical to
the non-aged form (\ref{msd}), for strong aging ($t_a\gg t$) the scaling with
the process time $t$ is linear and thus, deceivingly, identical to that of
normal Brownian diffusion. However, the presence of the power $t_a^{\alpha-1}$
is reminiscent of the anomaly $\alpha$ of the process. We note that the behavior
(\ref{5a}) and thus (\ref{6}) is identical to the result for the subdiffusive CTRW
\cite{johannes,eli_age} as well as aged heterogeneous diffusion processes with a
power-law form of the position dependent diffusivity \cite{hdp_age}.
 
In single particle tracking experiments \cite{orrit,saxton_rev,brauchle,xie} one
measures the time series $x(t)$ of the position of a labeled particle, which is
then typically evaluated in terms of the time averaged MSD. For an aged process
originally initiated at $t=0$ and measured from $t_a$ for the duration (measurement
time) $t$ this time averaged MSD is defined in the form \cite{johannes}
\begin{equation}
\label{2a}
\overline{\delta_a^2(\Delta)}=\frac{1}{t-\Delta}\int^{t+t_a-\Delta}_{t_a}
\Big[x(t'+\Delta)-x(t')\Big]^2dt',
\end{equation}
as a function of the lag time $\Delta$ and the aging time $t_a$. Averaging over
an ensemble of $N$ individual trajectories in the form
\begin{equation}
\left<\overline{\delta_a^2(\Delta)}\right>=\frac{1}{N}\sum_{i=1}^N\overline{\delta
_{a,i}^2(\Delta)},
\end{equation}
the structure function $\langle\left[x(t'+\Delta)-x(t')\right]^2\rangle$ in the
integral of expression (\ref{2a}) can be evaluated in terms of the covariance
(\ref{5a1}). The exact result reads
\begin{eqnarray}
\nonumber
\left<\overline{\delta_a^2(\Delta)}\right>&=&\frac{2K_{\alpha}^*}{(\alpha+1)(t-
\Delta)}\Big[(t+t_a)^{\alpha+1}-(t_a+\Delta)^{\alpha+1}\\
&&-(t+t_a-\Delta)^{\alpha+1}+t_a^{\alpha+1}\Big].
\label{3a}
\end{eqnarray}
In the absence of aging, we recover the known result \cite{fulinsky,thiel,sbm}
\begin{equation}
\label{tamsd_nonage}
\left<\overline{\delta^2(\Delta)}\right>\sim2K^*\frac{\Delta}{t^{1-\alpha}}.
\end{equation}
Its linear lag time dependence contrasts the power-law form of the ensemble
averaged MSD (\ref{msd}) and thus demonstrates that this process is weakly
non-ergodic in the sense of the disparity \cite{pt,bouchaud_web,pccp}
\begin{equation}
\left<\overline{\delta^2(\Delta)}\right>\neq\langle x^2(\Delta)\rangle.
\end{equation}
The equivalence and therefore ergodicity in the Boltzmann sense is only
restored in the Brownian case $\alpha=1$. In the presence of aging, expansion
of expression (\ref{3a}) in the limit $t,t_a\gg\Delta$ of short lag times yields
\begin{equation}
\label{4a}
\left<\overline{\delta_a^2(\Delta)}\right>\sim\Lambda_{\alpha}(t_a/t)\times
\left<\overline{\delta^2(\Delta)}\right>,
\end{equation}
in which we defined the so-called aging depression as
\begin{equation}
\label{depress}
\Lambda_{\alpha}(z)=(1+z)^{\alpha}-(z)^{\alpha}.
\end{equation}
In this experimentally relevant limit all the information on the age of the process
is thus contained in the aging depression $\Lambda_{\alpha}$, and the physically
important dependence on the lag time $\Delta$ factorizes, such that Eq.~(\ref{4a})
contains the non-aged form (\ref{tamsd_nonage}). Result (\ref{4a}) is identical to
the behavior of aged subdiffusive CTRW \cite{johannes} and heterogeneous diffusion
processes \cite{hdp}. In the limit $t_a\gg t$ of strong aging, the time
averaged MSD (\ref{3a}) remarkably reduces to the form
\begin{equation}
\label{6a}
\left<\overline{\delta_a^2(\Delta)}\right>=2\alpha K_{\alpha}^*t_a^{\alpha-1}
\Delta.
\end{equation}
In this limit, the time averaged MSD thus becomes equivalent to the aged ensemble
averaged MSD, $\langle\overline{\delta_a^2(\Delta)}\rangle=\langle x^2(\Delta)
\rangle_a$, as evidenced by comparison with result (\ref{5a}). In this limit,
that is, ergodicity is restored, as already observed for aged CTRW processes
\cite{johannes}.

\begin{figure}
\includegraphics[width=7.8cm]{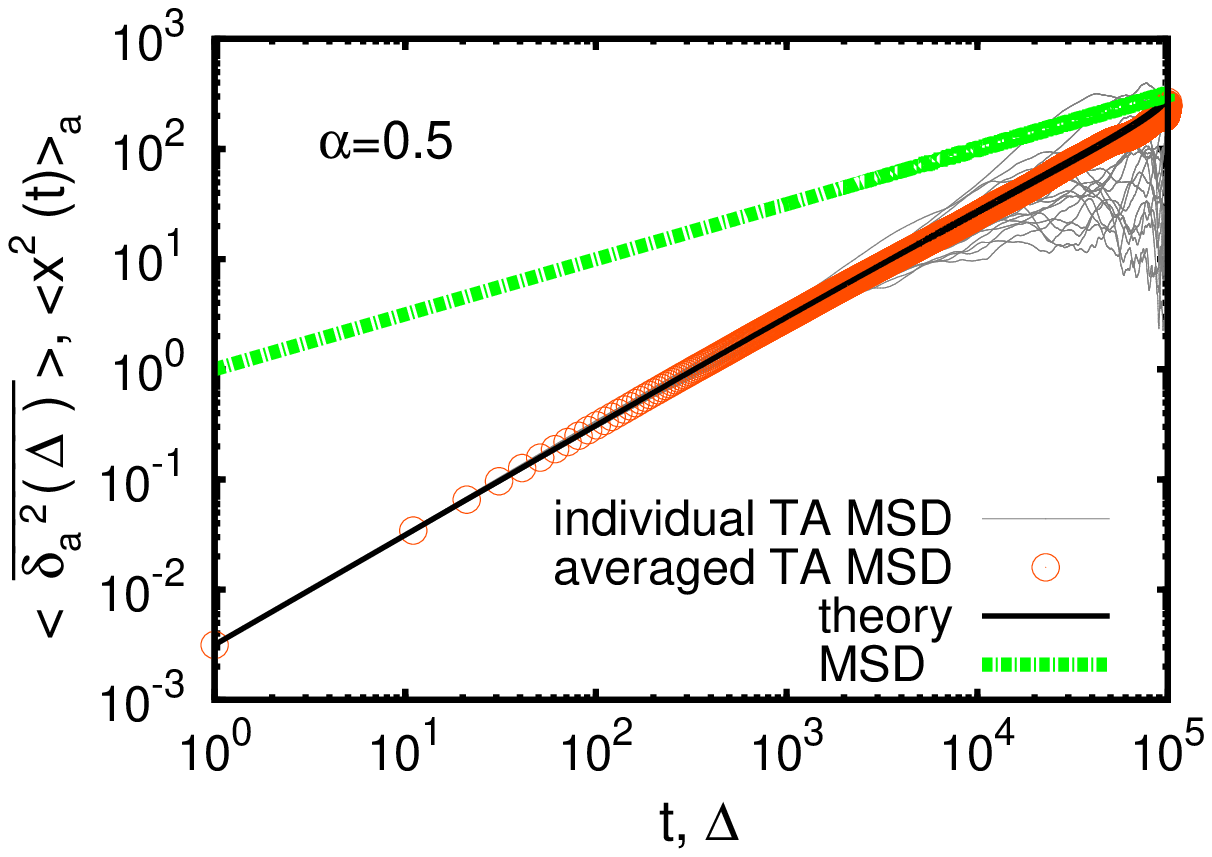}
\includegraphics[width=7.8cm]{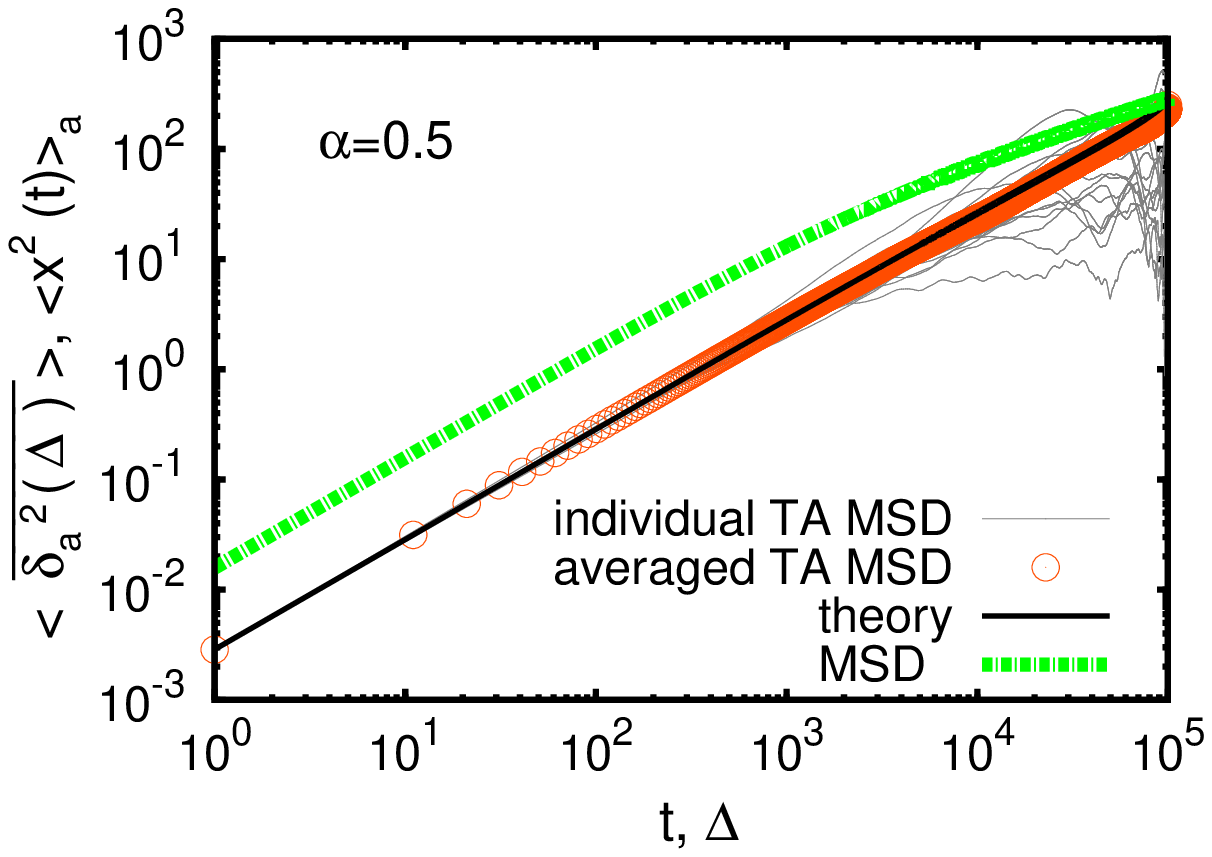}
\includegraphics[width=7.8cm]{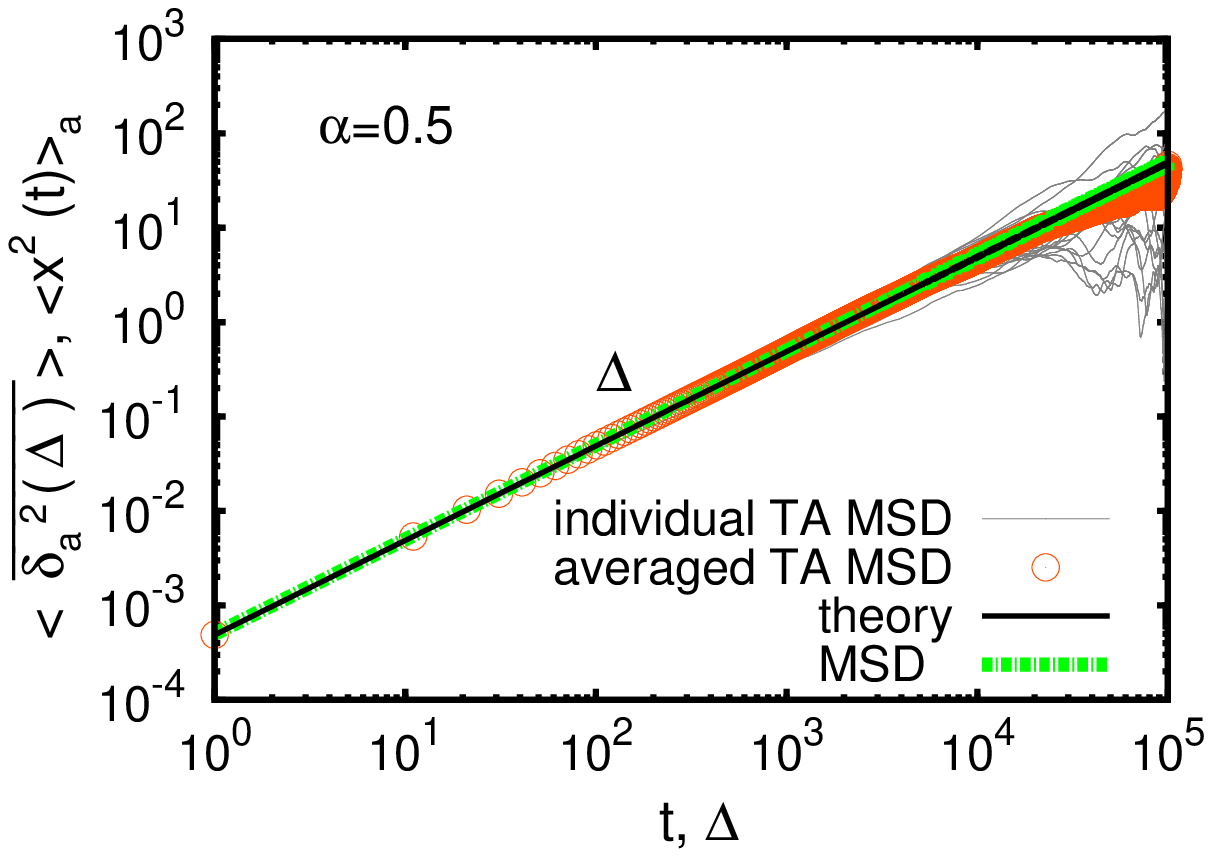}
\caption{Ensemble and time averaged MSD for SBM with $\alpha=1/2$. Thin
lines: time averaged MSD for 20 individual trajectories from simulations of
the SBM Langevin equation ($\ref{langevin}$) with trajectory length $t=10^5$.
Circles: averages over those 20 trajectories. Black thin line: theory result
(\ref{4a}). Thick green line: ensemble averaged MSD (\ref{5a}). Three different
aging times were considered (top to bottom): (a) non-aged case $t_a=0$, (b)
weak aging case $t_a =10^3$, and (c) strong aging case $t_a=10^6$. In all
simulations $K_{\alpha}^*=1/2$.}
\label{ergodic1}
\end{figure}

Figure \ref{ergodic1} shows the behavior of the ensemble and time averaged
MSD for unconfined SBM at different aging times in the subdiffusive case with
$\alpha =1/2$. The thin lines depict the simulations results for the time averaged
MSD for 20 individual trajectories. The first observation is that the amplitude
spread between these 20 time traces is fairly small. Note that the larger scatter
for longer lag times $\Delta$ is due to worsening statistics when $\Delta$
approaches the trace length $t$. The circles in Fig.~\ref{ergodic1} correspond
to the average over the 20 different results for the time averaged MSD. The
latter compare very nicely with the theoretical expectation (\ref{4a}). Finally,
the thick green line is the theoretical result (\ref{5a}) for the ensemble
averaged MSD. The detailed behavior in the three different aging regimes is
as follows:

(i) In the non-aged case ($t_a=0$, top panel of Fig.~\ref{ergodic1}) the power-law
growth $\langle x^2(t)\rangle\simeq t^{\alpha}$ of the MSD contrasts the linear
form $\langle\overline{\delta^2(\Delta)}\rangle\simeq\Delta$, this disparity being
at the heart of the weak ergodicity breaking \cite{fulinsky,thiel,sbm}.

(ii) In the weak aging case ($t_a=10^2$, middle panel of Fig.~\ref{ergodic1}) a
major change is visible in the behavior of the MSD, namely, we see the crossover
from the aging-dominated linear scaling $\langle x^2(t)\rangle\simeq t_a^{\alpha
-1}t$ to the anomalous scaling $\langle x^2(t)\rangle\simeq t^{\alpha}$, encoded in
Eqs.~(\ref{6}). The behavior of the time averaged MSD is
largely unchanged in comparison to case (i).

(iii) In the strong aging case ($t_a=10^6$, bottom panel of Fig.~\ref{ergodic1})
we see the apparent restoration of ergodicity: ensemble and time averaged MSDs
coincide, as given by Eq.~(\ref{6a}).

\begin{figure}
\includegraphics[width=8cm]{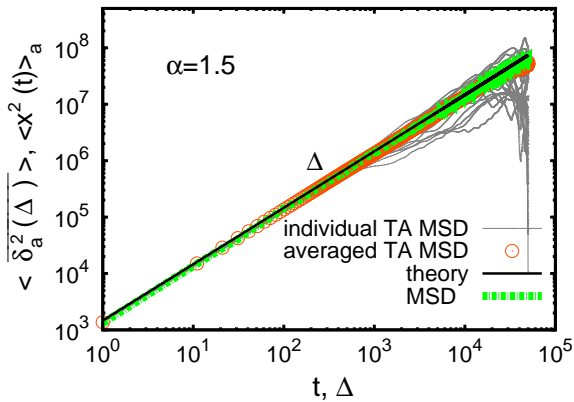}
\caption{Ensemble and time averaged MSD for SBM ($\ref{langevin}$) with $\alpha=
3/2$ in the strong aging case, $t_a=10^6$. The observation time is $t=10^5$. The
spread of the 20 single trajectory time averages is fairly small. As before, the
ensemble and time averaged MSDs coincide, apparently restoring ergodicity.}
\label{ergidic2}
\end{figure}

The convergence of the ensemble and time averaged MSDs in the strong aging case
for the superdiffusive case with $\alpha=3/2$ is nicely corroborated in
Fig.~\ref{ergidic2}.

\section{Ageing effect on confined SBM}
\label{sec:3}

In many cases an observed particle cannot be considered free during the
observation. Examples contain particles moving in confined space, for
instance, within the confines of living biological cells \cite{weber,he}.
Similarly, particles measured in optical tweezers setups experience a confining
Hookean force \cite{lene,lene1,warwick}. As a generic example for confined SBM
we consider the linear restoring force $-kx(t)$ with force constant $k$. The
corresponding stochastic equation for this confined SBM reads \cite{sbm}
\begin{equation}
\label{1b}
\frac{dx(t)}{dt}=-kx(t)+\sqrt{2\alpha K_{\alpha}^*t^{(\alpha-1)}}\times\xi(t),
\end{equation}
where, as before, $\xi(t)$ represents white Gaussian noise of zero mean. The
covariance in this confined case yields in the form \cite{sbm}
\begin{equation}
\label{A2}
\langle x(t_1)x(t_2)\rangle=2K_{\alpha}^*t_1^{\alpha}e^{-k(t_1+t_2)}M(\alpha,
\alpha+1,2kt_1)
\end{equation}
for $t_1<t_2$ in terms of the confluent hypergeometric function of the first kind,
also referred to as the Kummer function \cite{sbm,abramowitz}. Based on this
result we now present the ensemble and time averaged MSDs.

\subsection{Ensemble averaged MSD of confined SBM}

The ensemble averaged MSD for aging SBM, $\langle x^2(t)\rangle_a=\langle[x(t_a+t)
-x(t_a)]^2\rangle$ becomes
\begin{equation}
\label{2b}
\langle x^2(t)\rangle_a=2\mathscr{M}_1(t_a+t)+2\mathscr{M}_1(t_a)-4e^{-kt}
\mathscr{M}_1(t_a),
\end{equation}
where we used the abbreviation
\begin{equation}
\mathscr{M}_1(t)=K_{\alpha}^*t^{\alpha}\exp(-2kt)M(\alpha,\alpha+1,2kt).
\end{equation}
In the limit $k\rightarrow 0$ of vanishing confinement, Eq.~(\ref{5a}) for free
SBM is readily recovered from the property $M(\alpha,\alpha+1,0)=1$.

\begin{figure*}
\includegraphics[width=12cm]{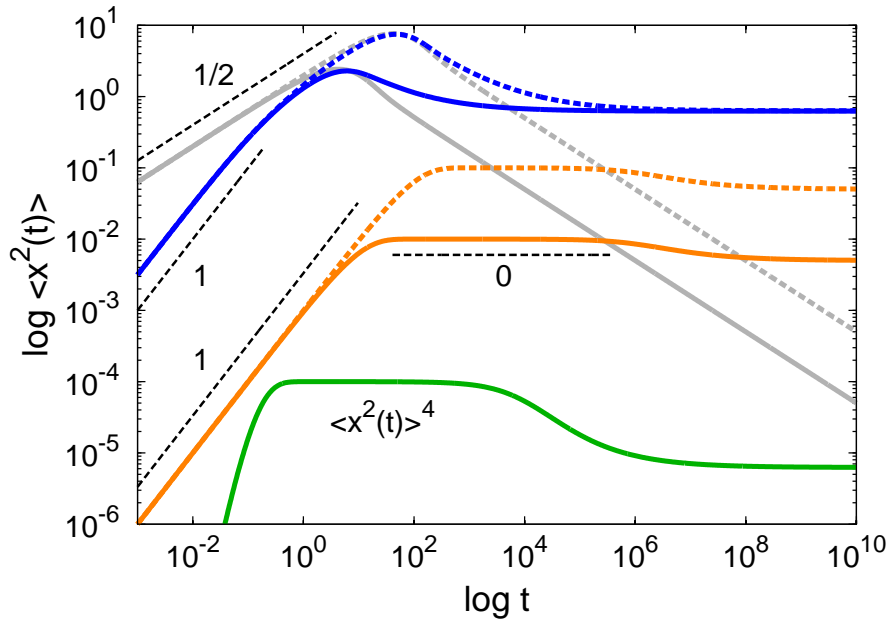}
\caption{Ensemble averaged MSD $\langle x^2(t)\rangle$ for confined aging SBM
in the subdiffusive case with $\alpha=0.5$
at different aging times: (i) non-aged ($t_a=0$) denoted by the grey lines;
(ii) weakly aged ($t_a=0.1$) denoted by the blue lines; and (iii) strongly
aged ($t_a=10^6$) denoted by the orange lines. In all cases, the full lines
correspond to the force constant $k=0.1$, while the dashed lines stand for
$k=0.01$. The green line at the bottom of the graph is a blowup ($\langle x^2(t)
\rangle^4$ of the case $t_a=10^6$ and $k=0.1$) in which the crossover between
the two plateaux is better visible.}
\label{msd_ens}
\end{figure*}

\begin{figure*}
\includegraphics[width=12cm]{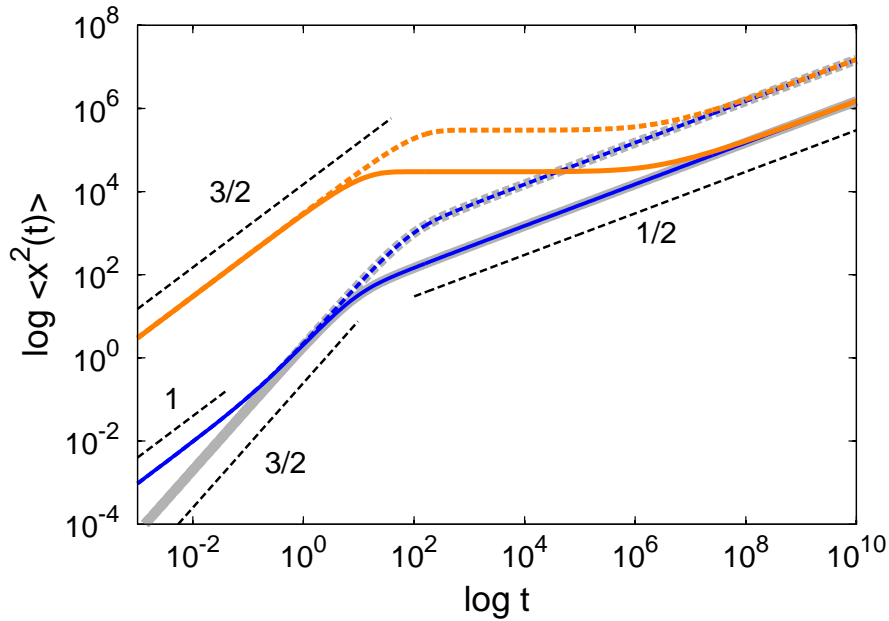}
\caption{Ensemble averaged MSD $\langle x^2(t)\rangle$ for confined aging SBM
in the superdiffusive case with $\alpha=1.5$
at different aging times: (i) non-aged ($t_a=0$) denoted by the grey lines;
(ii) weakly aged ($t_a=0.1$) denoted by the blue lines; and (iii) strongly
aged ($t_a=10^6$) denoted by the orange lines. In all cases, the full lines
correspond to the force constant $k=0.1$, while the dashed lines stand for
$k=0.01$. In all cases the terminal scaling $\simeq t^{\alpha-1}$ is reached.}
\label{msd_ens1}
\end{figure*}

We now discuss the result (\ref{2b}) in the three limits of the non-aged, weakly
aged, and strongly aged processes. The analysis reveals a rich behavior depending
on the values of the aging time $t_a$ and the anomalous diffusion exponent $\alpha$.
For sub- and superdiffusion, respectively, the various crossovers are displayed
in Figs.~\ref{msd_ens} and \ref{msd_ens1}.

(i) In the absence of aging ($t_a=0$), we get back to the result
\begin{equation}
\langle x^2(t)\rangle=2\mathscr{M}_1(t)
\end{equation}
reported in Ref.~\cite{sbm}. For $t\ll1/k$ this reduces to the non-aged free
SBM result (\ref{msd}), while in the long time limit $t\gg1/k$ we use the
expansion
\begin{equation}
\label{A3}
M(\alpha,\alpha+1,z)\sim\alpha\frac{\exp(z)}{z}
\end{equation}
of the Kummer function to obtain \cite{sbm}
\begin{equation}
\langle x(t)^2\rangle\sim\frac{\alpha K_{\alpha}^*}{k}t^{\alpha-1}.
\end{equation}
This result underlines the inherently non-stationary character of SBM: for
subdiffusion the MSD $\langle x(t)^2\rangle$ progressively decays, while for
superdiffusion it increases. This property reflects the time dependence of
the temperature encoded in the diffusivity (\ref{time_diff}) \cite{sbm}. This
non-aged behavior is shown in Figs.~\ref{msd_ens} and \ref{msd_ens1} as the
grey lines for two different strengths $k$ of the external confining potential.
How does aging modify this behavior?

(ii) We first consider the case $t_a\ll1/k$. If in addition $t\ll1/k$, this is but
the above result (\ref{5a}) for free aging SBM. However, care needs to be
taken when $t\gg1/k$. From Eq.~(\ref{2b}), we find that the first two terms
(the third one is exponentially small in $t$ and can be neglected) lead to the
asymptotic behavior
\begin{equation}
\langle x^2(t)\rangle_a\sim\frac{\alpha K_{\alpha}^*}{k}t^{\alpha-1}+2K_{\alpha}^*
t_a^{\alpha}.
\end{equation}
This implies that for subdiffusion $(0<\alpha<1$) the first term tends to zero
and the leading behavior is the plateau
\begin{equation}
\label{weak_plateau}
\langle x^2(t)\rangle_a\sim2K_{\alpha}^*t_a^{\alpha}.
\end{equation}
Even for very weak aging, the ensemble averaged MSD $\langle x^2(t)\rangle_a$
becomes $t_a$-dependent. When experimental data are evaluated and the exact
equivalence $t_a=0$ is not guaranteed, the erroneous conclusion could be drawn
that the process is stationary. Note, however, that result (\ref{weak_plateau})
is independent of the strength $k$ of the confining potential and only depends
on the diffusion coefficient $K_{\alpha}^*$ and the aging time $t_a$, mirroring
the fact that this term stems from the initial free motion during the aging period.
Conversely, for superdiffusion ($\alpha>1$) the leading order term indeed shows
the growth
\begin{equation}
\label{weak_growth}
\langle x^2(t)\rangle_a\sim\frac{\alpha K_{\alpha}^*}{k}t^{\alpha-1}
\end{equation}
of the ensemble averaged MSD. The weakly aged behavior is shown in
Figs.~\ref{msd_ens} and \ref{msd_ens1} as the blue lines.

(iii) With the asymptotic expansion (\ref{A3}) of the Kummer function, we find
that in the strong aging regime $t_a\gg1/k$ the ensemble averaged MSD becomes
\begin{equation}
\label{strong}
\langle x^2(t)\rangle_a\sim\alpha k^{-1}K_{\alpha}^*\left[(t_a+t)^{\alpha-1}+
t_a^{\alpha-1}\left(1-2e^{-kt}\right)\right].
\end{equation}
At short times $t\ll1/k$, this leads us back to the unconfined result $\langle x^2
(t)\rangle_a\sim2\alpha K_{\alpha}^*t_a^{\alpha-1}t$ of Eq.~(\ref{6}). At long
time $t\gg1/k$, however, we have to distinguish two different regimes. First,
for $t_a\gg t\gg1/k$ we obtain the plateau
\begin{equation}
\label{strong_strong}
\langle x^2(t)\rangle_a\sim\frac{2\alpha K_{\alpha}^*}{k}t_a^{\alpha-1},
\end{equation}
which differs from the above result (\ref{weak_growth}) by the factor of two.
Second, for $t\gg t_a\gg1/k$ the leading order according to Eq.~(\ref{strong})
again differs between sub- and superdiffusive motion. For $0<\alpha<1$ the
plateau
\begin{equation}
\langle x^2(t)\rangle_a\sim\frac{\alpha K_{\alpha}^*}{k}t_a^{\alpha-1}
\end{equation}
emerges. Note, however, that in comparison to Eq.~(\ref{strong_strong}) we now
have half the amplitude. In the superdiffusive case $\alpha>1$ we recover
result (\ref{weak_growth}). This intricate behavior is shown in
Figs.~\ref{msd_ens} and \ref{msd_ens1} as the orange lines. In Fig.~\ref{msd_ens}
we pronounce the crossover between the two plateaux by plotting the fourth
power of the ensemble MSD as the green line.

\subsection{Time averaged MSD of confined SBM}

The time averaged MSD for confined SBM can be derived by substituting the above
covariance (\ref{A2}) into the integral (\ref{2a}). By help of the relation
\cite{prudnikov}
\begin{eqnarray}
\nonumber
\int^xy^{\alpha}e^{-y}M(\alpha,1+\alpha,y)dy&=&\\
&&\hspace*{-3.2cm}\frac{1}{1+\alpha}x^{1+\alpha}e^{-x}
M(1+\alpha,2+\alpha,x)
\end{eqnarray}
this procedure yields the general result
\begin{eqnarray}
\nonumber
\left<\overline{\delta_a^2(\Delta)}\right>&=&\frac{2K_{\alpha}^*}{(t-\Delta)(1+
\alpha)}\Big[\mathscr{M}_2(t+t_a)-\mathscr{M}_2(t_a+\Delta)\\
\nonumber
&&+\mathscr{M}_2(t+t_a-\Delta)-\mathscr{M}_2(t_a)\\
&&-2e^{-k \Delta}\Big(\mathscr{M}_2(t+t_a-\Delta)-\mathscr{M}_2(t_a)\Big)\Big],
\label{A1}
\end{eqnarray}
where we used the abbreviation
\begin{equation}
\mathscr{M}_2(t)=t^{1+\alpha}e^{-2kt}M(1+\alpha,2+\alpha,2kt).
\end{equation}

(i)
In the limit $k\rightarrow 0$ we recover the result (\ref{3a}) of unconfined aging
SBM, while the complete absence of aging restores the result from Ref.~\cite{sbm}.
In the presence of confinement, we distinguish the following regimes.

(ii) We now consider the case when the aging time is short compared to the relaxation
time of the system, $t_a\ll1/k$. From the general expression (\ref{A1}) we then
find the following behaviors: (a) when in addition the lag time is short ($t\gg1/k
\gg\Delta\gtrsim t_a$) we recover the non-aged result (\ref{tamsd_nonage}) with its
linear scaling in the lag time $\Delta$. (b) When the lag time is long ($t\gg
\Delta\gg1/k\gg t_a$), we find the plateau
\begin{equation}
\label{plateau}
\left<\overline{\delta^2_a(\Delta)}\right>\sim\frac{2K_{\alpha}^*}{k}t^{\alpha-1}
\end{equation}
known from the non-aged case \cite{pccp}. (c) Finally, when the lag time
approaches the length $t$ of the time series, the time averaged MSD
\begin{equation}
\left<\overline{\delta^2_a(\Delta)}\right>\sim\frac{\alpha K_{\alpha}^*}{k}
t^{\alpha-1}
\end{equation}
becomes equivalent to the ensemble averaged MSD, Eq.~(\ref{weak_growth}). In
contrast to the ensemble averaged MSD, we thus find that the time averaged MSD
is not affected by short aging times as compared to the relaxation time scale,
$t_a\ll1/k$.

\begin{figure}
\includegraphics[width=7.8cm]{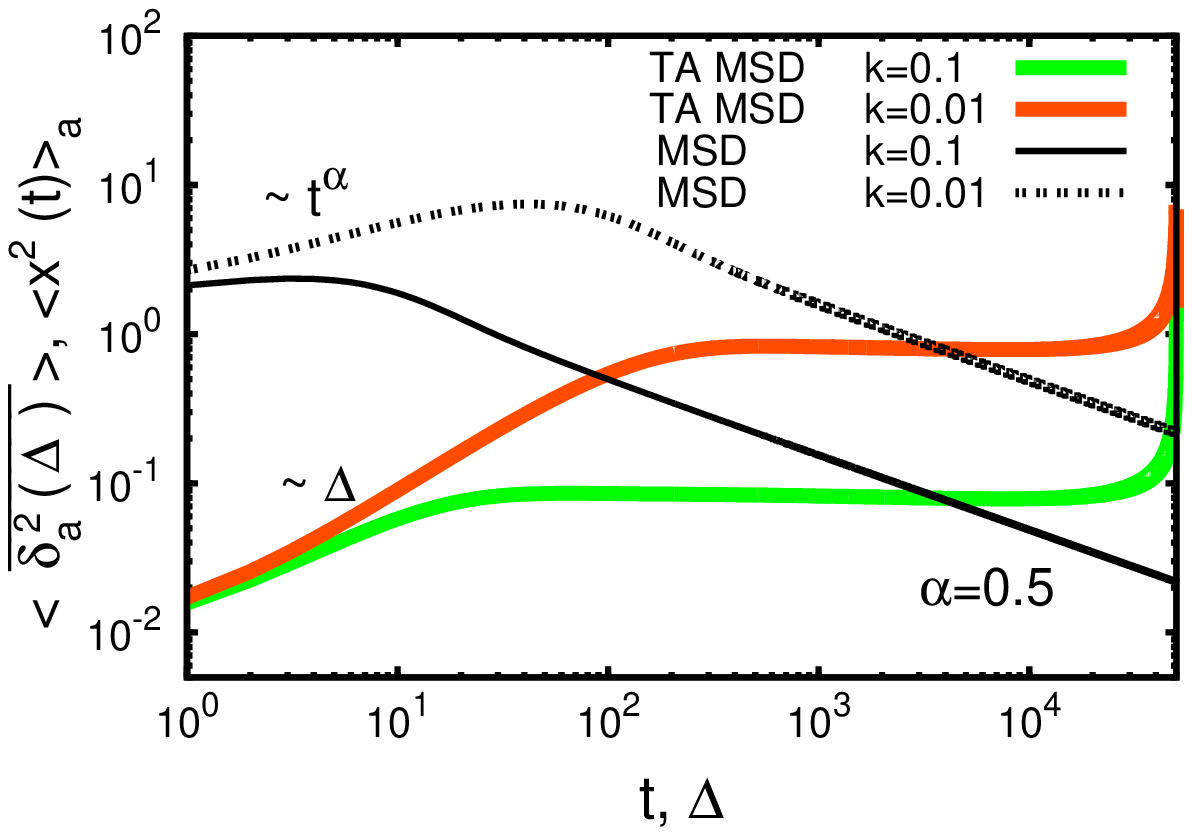}
\includegraphics[width=7.8cm]{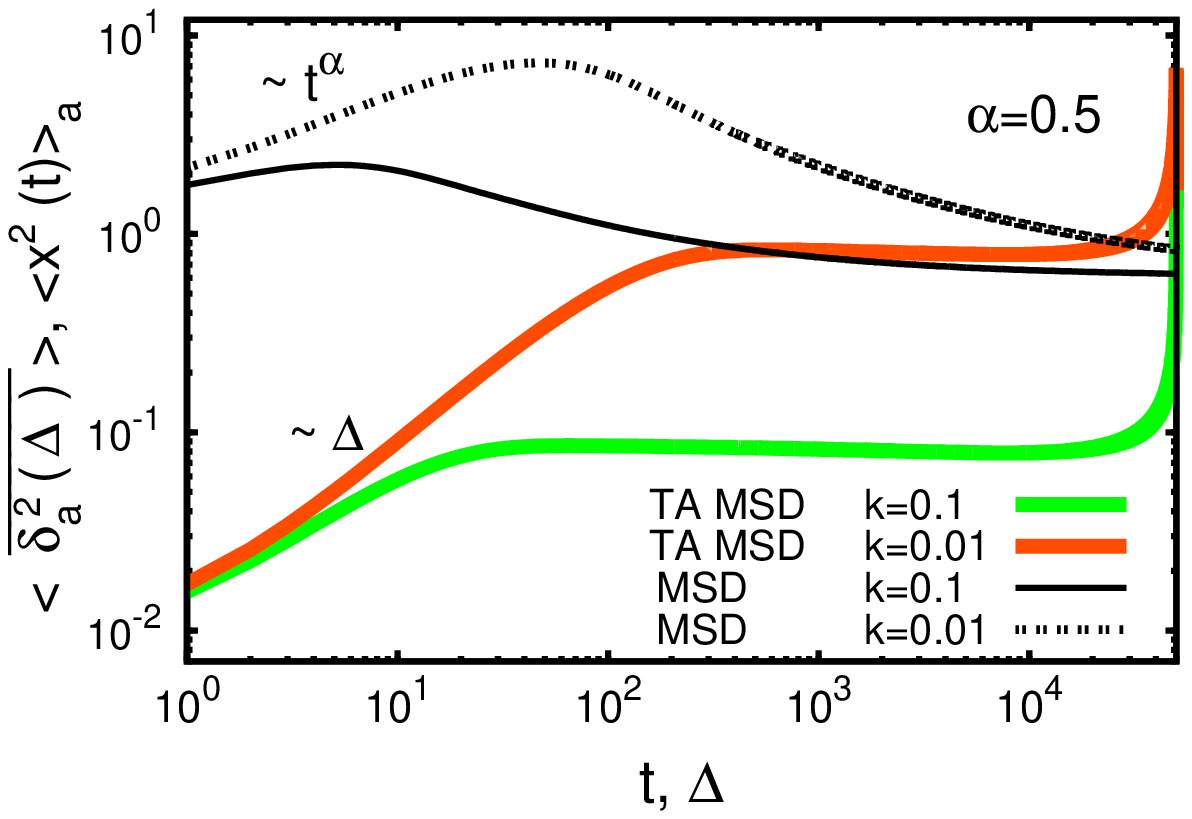}
\includegraphics[width=7.8cm]{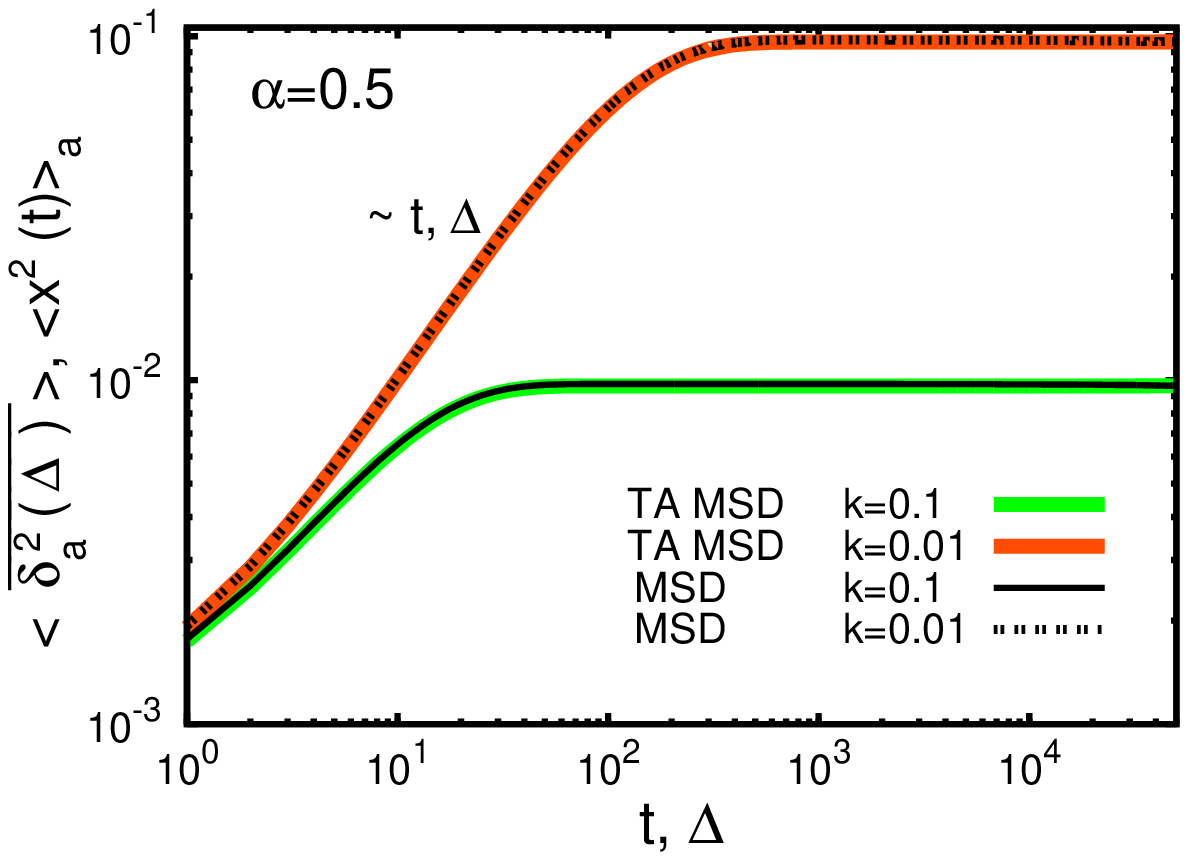}
\caption{Ensemble and time averaged MSD for confined SBM for $\alpha=1/2$. From
top to bottom, the panels represent the non-aged ($t_a=0$) case, the case of
weak aging ($t_a=10^{-1}$), and the case of strong aging ($t_a=10^6$), where
the observation time is chosen as $t=5\times10^4$. The lines represent
Eqs.~(\ref{2b}) and (\ref{A1}). The force constants $k$ are indicated in the
panels. Note that the time averaged MSD indeed converges to the ensemble MSD
in the limit $\Delta\to t$, compare the discussion in Ref.~\cite{sbm}.}
\label{Confined_SBM}
\end{figure}

(iii)
The second, more interesting case corresponds to long aging times compared to the
relaxation time scale, $t_a\gg1/k$. When also $t\gg1/k$, the result is independent
of the specific magnitude of the lag time. From the general expression (\ref{A1})
by help of relation (\ref{A3}) we obtain
\begin{eqnarray}
\nonumber
\left<\overline{\delta_a^2(\Delta)}\right>&\sim&\frac{K_{\alpha}^*}{k(t-\Delta)}
\Big[(t+t_a)^{\alpha}-(\Delta+t_a)^{\alpha}\\
&&+(1-2e^{-k\Delta})[(t+t_a-\Delta)^{\alpha}-t_a^{\alpha}]\Big].
\label{A4}
\end{eqnarray}
If we now consider the regime in which the lag time is short, $t,t_a\gg1/k\gg
\Delta$, we obtain result (\ref{4a}) with the aging depression (\ref{depress})
from unconfined aging SBM. In the opposite limit $t,t_a\gg\Delta\gg1/k$ when
the lag time is long compared to the relaxation time, we find
\begin{equation}
\left<\overline{\delta_a^2(\Delta)}\right>\sim\Lambda_a(t_a/t)\left<\overline{
\delta^2(\Delta)}\right>,
\end{equation}
where $\left<\overline{\delta^2(\Delta)}\right>$ is equal to expression
(\ref{plateau}) and $\Lambda_a(z)$ is again the aging depression (\ref{depress}).
In the strong aging limit $t,t_a\gg\Delta\gg1/k$, that is, the aged time
averaged MSD is generally given by $\left<\overline{\delta_a^2(\Delta)}\right>
\sim\Lambda_a(t_a/t)\left<\overline{\delta^2(\Delta)}\right>$ for any lag time.
Similar to subdiffusive CTRW processes \cite{johannes}, the occurrence of the
factor $\Lambda_a$ appears like a general feature for the aging dynamics of SBM.

\begin{figure}
\includegraphics[width=8cm]{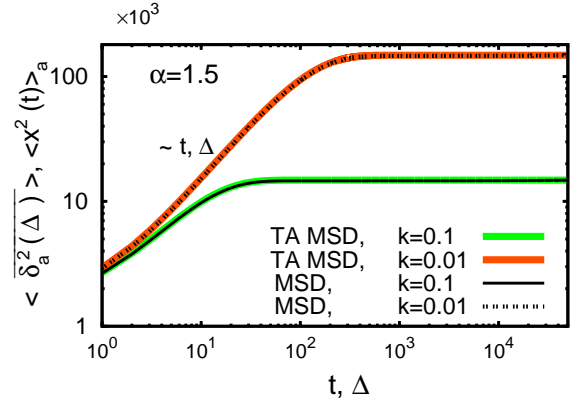}
\caption{Ensemble and time averaged MSD for confined SBM with $\alpha=3/2$ in
the strong aging case with $t_a=10^6$. The observation time is $t=5\times10^4$.}
\label{Confined SBM 2}
\end{figure}

\begin{figure*}
\centering
\includegraphics[width=6.75cm]{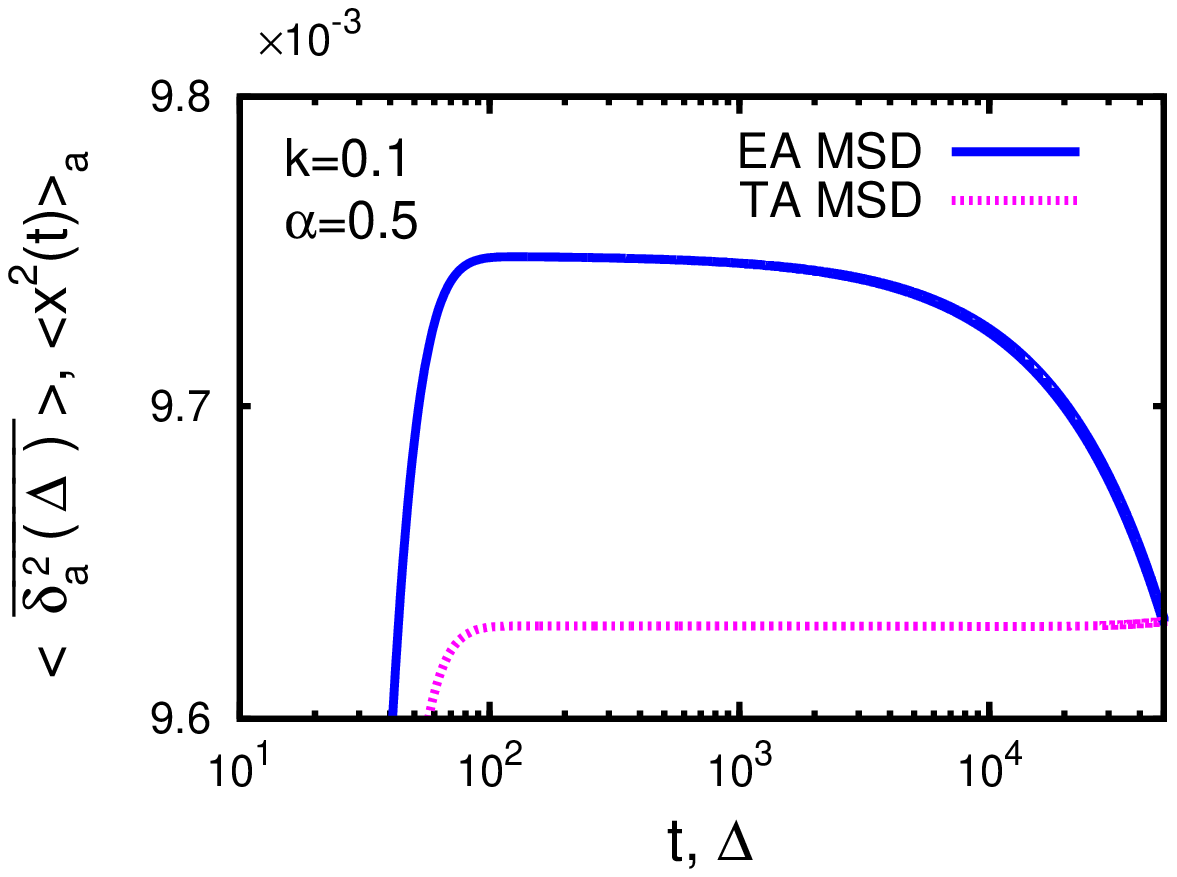}
\includegraphics[width=6.75cm]{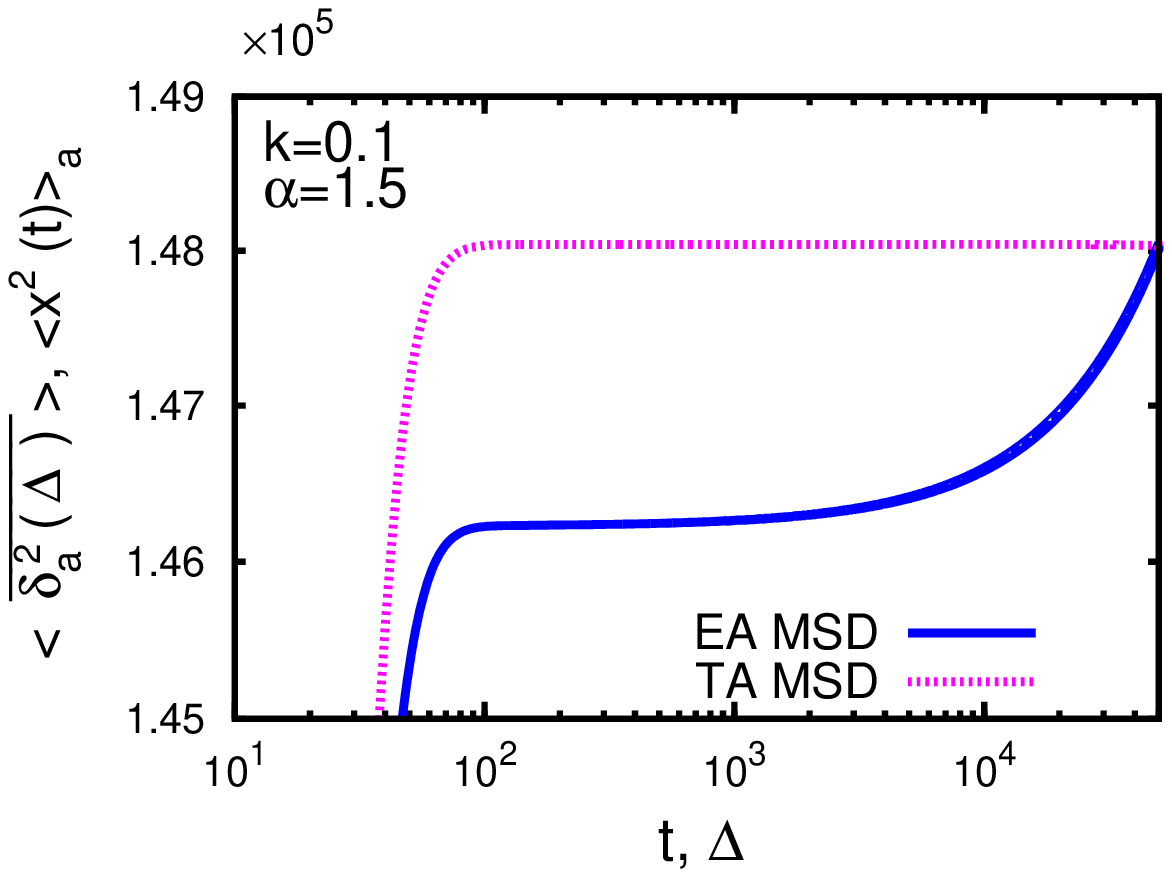}
\includegraphics[width=6.75cm]{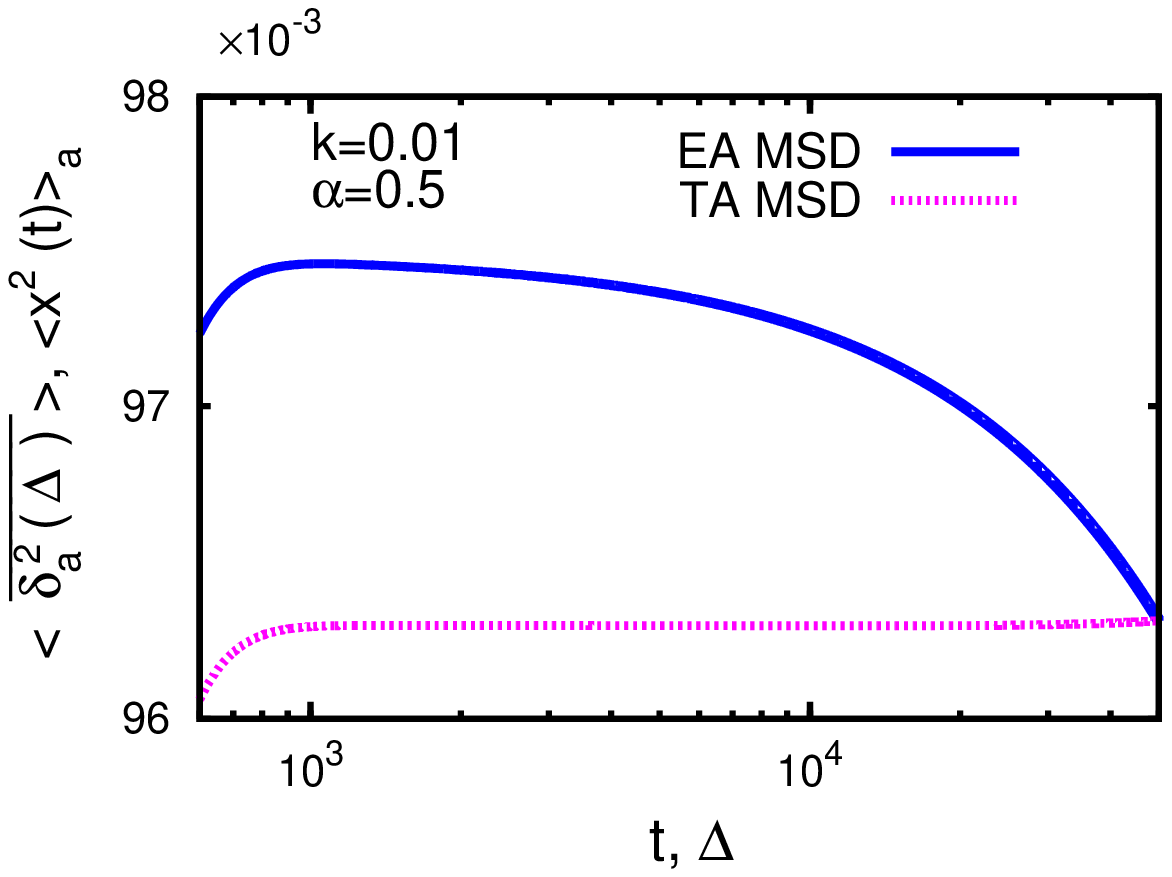}
\includegraphics[width=6.75cm]{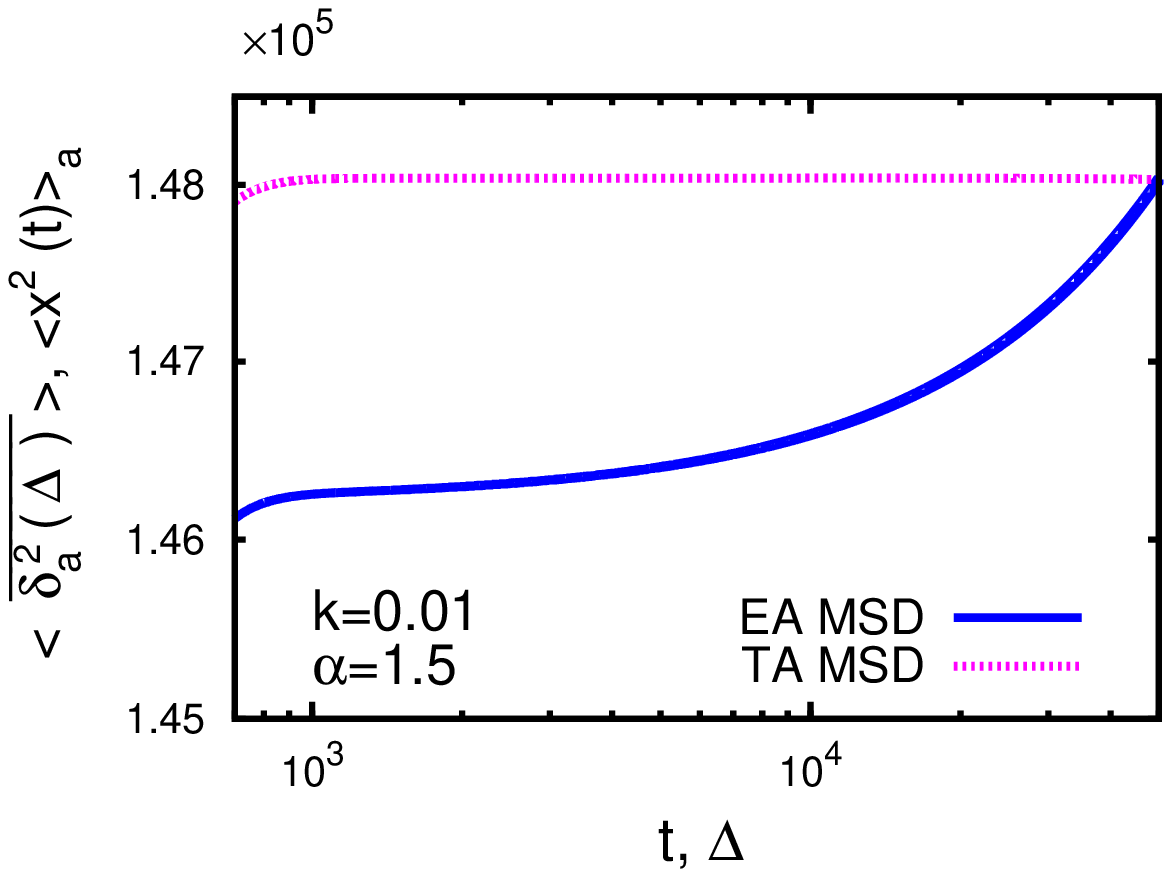}
\caption{Full behavior of aging confined SBM for $t_a=10^6$ with $\alpha=1/2$
(left) and $\alpha=3/2$ (right), demonstrating the convergence of the time averaged
MSD to the ensemble averaged MSD in the limit $\Delta\rightarrow t$, shown for two
different potential strengths, as indicated in the panels.}
\label{discrepancy}
\end{figure*}

Figure \ref{Confined_SBM} shows the behavior of the ensemble and time averaged
MSD for confined SBM at different degrees of aging. The graphs represent the full
behavior according to Eqs.~(\ref{2b}) and (\ref{A1}). In the absence of aging,
the initial linear growth $\langle\overline{\delta_a^2(\Delta)}\rangle\simeq
\Delta$ of the time averaged MSD crosses over to an apparent plateau, contrasting
the functional behavior of the ensemble average: at short times, we observe the
power-law growth $\langle x^2(t)\rangle_a\simeq t^{\alpha}$ of unconfined SBM,
while after engaging with the confining potential, the monotonic decrease $\langle
x^2(t)\rangle_a\simeq t^{\alpha-1}$ reflects the temporal decay of
the noise strength (i.e., the temperature) \cite{sbm}.
When aging effects come into play, the ensemble averaged MSD displays notable
differences. In the case of weak aging displayed in the middle panel of
Fig.~\ref{Confined_SBM} deviations from the power-law decay of $\langle x^2(t)
\rangle_a$ become apparent for longer times, $t\gg t_a\gg1/k$. Eventually the
convergence to a common value independent of the force constants is observed,
as predicted by Eq.~(\ref{weak_plateau}). Finally, in the strong aging limit, the
ensemble and time averaged MSD are equivalent and ergodicity is seemingly restored:
$\langle\overline{\delta_a^2(\Delta)}\rangle=\langle x^2(\Delta)\rangle_a$, as can
be witnessed in the bottom panel of Fig.~\ref{Confined_SBM}. The apparent
equivalence of ensemble and time averaged MSDs in the strong aging limit is also
proven in the superdiffusive case for $\alpha=3/2$ in Fig.~\ref{Confined SBM 2}.
The slight discrepancy remaining between time and ensemble averaged MSD in the
latter strong (but finite) aging case is shown in Fig.~(\ref{discrepancy}). This
Figure also demonstrates the convergence of ensemble and time averaged MSDs.

\section{ First Passage Time Density}
\label{sec:4}

Apart from the MSD the first passage behavior is a signature quantity of a
stochastic process. We here study how aging changes the first passage
statistic of SBM in the semi-infinite domain. The probability density function
(PDF) of first passage is found by solving the SBM diffusion equation with the
time dependent coefficient $\mathscr{K}(t)$,
\begin{equation}
\label{1c}
\frac{\partial}{\partial t}P(x,t)=\mathscr{K}(t)\frac{\partial^2}{\partial x^2}
P(x,t),
\end{equation}
however, with the aged initial condition
\begin{equation}
\label{2c}
P_0(x,t_a)=\frac{1}{\sqrt{4\pi K_{\alpha}^*t_a^{\alpha}}}e^{-x^2/(4K_{\alpha}^*
t_a^{\alpha})}.
\end{equation}
This aged initial condition emerges from a $\delta(x)$ peak for a system initialized
some aging time $t_a$ before. In this setup $t$ measures the time span from the
aged initial condition (\ref{2c}). To obtain the first passage PDF for the
semi-infinite domain we solve the SBM diffusion equation (\ref{1c}) for unconfined
motion with the aged initial condition (\ref{2c}) and then use the method of images.
For the PDF of the aged process we obtain
\begin{equation}
\label{3c}
P(x,t)=\frac{1}{\sqrt{4\pi K_{\alpha}^*(t_a^{\alpha}+t^{\alpha})}}e^{-x^2/4K_
{\alpha}^*(t_a^{\alpha}+t^{\alpha})}.
\end{equation}

\begin{figure}
\includegraphics[width=8cm]{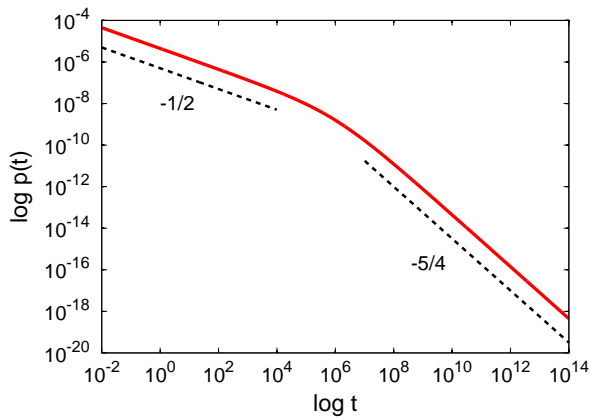}
\caption{First passage time density $\wp(t)$ for $\alpha=1/2$, with $x_0=1$, and
aging time $t_a=10^6$. As shown by the dashed line, the crossover between the
aging dominated slope $-1/2$ to the slope $-5/4$ is distinct.}
\label{FPTD}
\end{figure}

In the presence of an absorbing boundary at the origin, the survival probability
for a process initiated originally in $x_0>0$ is therefore given by
\begin{equation}
\label{4c}
\mathscr{S}(t)=\int_0^{\infty}\left[P(x-x_0,t)-P(x+x_0,t)\right]dx.
\end{equation}
Substituting the aged PDF (\ref{2c}) into this expression yields
\begin{equation}
\label{18}
\mathscr{S}(t)=\mathrm{erf}\left(\frac{x_0}{\sqrt{4K_{\alpha}^*(t_a^{\alpha}+t^{
\alpha})}}\right)
\end{equation}
in terms of the error function. The first passage PDF follows from the relation
$\wp(t)=-d\mathscr{S}(t)/dt$,
\begin{equation}
\label{5c}
\wp(t)=\frac{\alpha x_0t^{\alpha-1}}{\sqrt{4\pi K_{\alpha}^*\left(t_a^{\alpha}
+t^{\alpha}\right)^3}}\exp\left(-\frac{x_0^2}{4K_{\alpha}^*(t_a^{\alpha}+t^{
\alpha})}\right).
\end{equation}
For $\alpha=1$ (Brownian motion) and in the absence of aging ($t_a=0$) we recover
the well known L{\'e}vy-Smirnov distribution.
Result (\ref{5c}) exhibits a crossover relative to the aging time,
\begin{equation}
\wp\simeq\frac{\alpha x_0}{\sqrt{4\pi K_{\alpha}^*}}\times\left\{\begin{array}{ll}
t_a^{-3\alpha/2}t^{\alpha-1}, & t_a\gg t,(x_0^2/[4K_{\alpha}^*])^{1/\alpha}\\[0.4cm]
t^{-1-\alpha/2}, & t\gg t_a,(x_0^2/[4K_{\alpha}^*])^{1/\alpha}
\end{array}\right..
\end{equation}
In the strong aging limit the scaling exponent is $-(1-\alpha)$, and we observe the
explicit presence of the aging time $t_a$ with exponent $\alpha$. For weak aging,
the scaling exponent of $t$ is $-(1+\alpha/2)$, as known from subdiffusive CTRW
processes. However, the detailed crossover behavior is different, compare
Ref.~\cite{henning}. We also note that for fractional Brownian motion the anomalous
diffusion exponent $\alpha$ enters oppositely \cite{fbm_fpt,lim}. Fig.~(\ref{FPTD})
shows the crossover of the first passage density for aging SBM.

\section{Conclusions}
\label{sec:5}

SBM is possibly the simplest anomalous diffusion model, and it is therefore
widely used in literature. Despite its apparent simplicity SBM exhibits weak
ergodicity breaking in the sense that we observe a distinct disparity between
the ensemble and time averaged MSDs of this process \cite{pccp,thiel,fulinsky,
sbm}. It is therefore a natural question to explore the aging effects of SBM,
i.e., the explicit dependence of physical observables on the time span $t_a$
between the original system preparation. We here showed how the ensemble and
time averaged MSDs depend on $t_a$ for both the unconfined and confined cases.

For unconfined aging SBM we obtained the exact dependence on the aging time $t_a$
and observed a striking similarity to both subdiffusive CTRW with scale-free
waiting time distributions and heterogeneous diffusion processes with power-law
position dependence of the diffusivity. In particular, for short lag times the
time averaged MSD factorizes into the non-aged expression and the aging
depression $\Lambda_a$, which indeed has the same functional form as for the
subdiffusive CTRW and the heterogeneous diffusion process. In the limit of
strong aging, we also showed that ergodicity is seemingly restored and the
disparity between ensemble and time averages becomes increasingly marginal.

Confined aging SBM, in contrast, is qualitatively a quite unique process. Due
to the time dependence of the diffusivity $\mathscr{K}(t)$ there is
no thermal plateau for the ensemble averaged MSD. Instead this quantity is
continuously decaying (subdiffusion) or increasing (superdiffusion). As shown
here the functional behavior of confined aging SBM is remarkably rich.
Concurrently,
the time averaged MSD exhibits an intermediate plateau. In the presence of aging,
we observe a deviation from the non-aged power-law behavior of the ensemble
averaged MSD at longer times and a universal convergence to a plateau value.
For strong aging, we again observe the convergence of ensemble and time
averaged MSDs. We note that the behavior of confined CTRW is opposite: the time
averaged MSD exhibits a power-law growth with exponent $1-\alpha$, while the
ensemble averaged MSD converges to the thermal plateau value \cite{stas}.

In addition to these quantities we considered the first passage time density in
the semi-infinite domain. In contrast to the non-aging fractional Brownian motion
we found a crossover between two characteristic scaling laws depending on the
competition between aging and process time, $t_a$ and $t$.

When using SBM as a stochastic model cognizance should be taken of the fact
that it is a highly non-stationary process. The time dependence of its
diffusivity corresponds to a time dependent temperature (noise strength),
and is therefore physically meaningless as description for a system coupled
to a thermostat. There exist, however, cases, in which SBM may turn out to be
a physically meaningful approach. For instance, it was demonstrated that SBM
provides a useful mean field description for the motion of a tagged particle
in a granular gas with a sub-unity restitution coefficient in the homogeneous
cooling phase \cite{anna}.

A number of aging features are quite similar between subdiffusive CTRWs
\cite{johannes}, heterogeneous diffusion processes \cite{hdp_age}, and aging
SBM, as shown here. While fractional Brownian motion is ergodic, transient
deviations from ergodicity and transient aging occur under confinement
\cite{jochen,lene1}. To reliably distinguish these models from another, it
is therefore imperative to employ other diagnostic stochastic quantities
with characteristic behaviors for the respective processes \cite{pccp,vincent,
pvar}.

\end{document}